\documentclass{article}
\usepackage[a4paper, total={6in, 8in}]{geometry}

\usepackage{cite}
\usepackage{amsmath,amssymb,amsfonts}
\usepackage{algorithm}
\usepackage{algpseudocode}
\usepackage{multicol}
\usepackage{graphicx}
\usepackage{textcomp}
\usepackage{xcolor}

\usepackage{subcaption}

\DeclareMathOperator*{\argmin}{arg\,min}

\def\BibTeX{{\rm B\kern-.05em{\sc i\kern-.025em b}\kern-.08em
    T\kern-.1667em\lower.7ex\hbox{E}\kern-.125emX}}
\begin{document}

\title{Non-parametric B-spline decoupling of \\multivariate functions}

\author{Joppe De Jonghe \\
\textit{Dept. of Computer Science} \\
\textit{NUMA, KU Leuven}\\
Geel, Belgium \\
joppe.dejonghe@kuleuven.be
\and
Mariya Ishteva \\
\textit{Dept. of Computer Science} \\
\textit{NUMA, KU Leuven}\\
Geel, Belgium \\
mariya.ishteva@kuleuven.be}

\date{}

\maketitle

\begin{abstract}
Many scientific fields and applications require compact representations of multivariate functions. For this problem, decoupling methods are powerful techniques for representing the multivariate functions as a combination of linear transformations and nonlinear univariate functions. This work introduces an efficient decoupling algorithm that leverages the use of B-splines to allow a non-parametric estimation of the decoupling's internal functions. The use of B-splines alleviates the problem of choosing an appropriate basis, as in parametric methods, but still allows an intuitive way to tweak the flexibility of the estimated functions. Besides the non-parametric property, the use of B-spline representations allows for easy integration of nonnegativity or monotonicity constraints on the function shapes, which is not possible for the currently available (non-)parametric decoupling methods. The proposed algorithm is illustrated on synthetic examples that highlight the flexibility of the B-spline representation and the ease with which a monotonicity constraint can be added. The examples also show that if monotonic functions are required, enforcing the constraint is necessary.
\\
\textbf{Keywords: }Tensor, Decomposition, Decoupling, B-spline
\end{abstract}

\section{Introduction}
\begin{figure}
    \centering
    \includegraphics[width=0.75\linewidth]{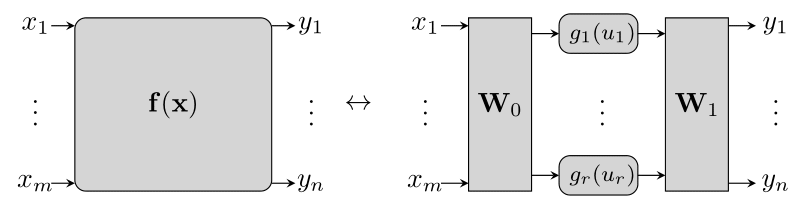}
    \caption{Decoupling of a multivariate function $\mathbf{f}(\mathbf{x})$ (left) into the model $\mathbf{f}(\mathbf{x}) = \mathbf{W}_1 \mathbf{g}(\mathbf{W}_0\mathbf{x})$ (right) constitutes a linear transformation of the input by $\mathbf{W}_0$, followed by branches of univariate functions and a final linear transformation by $\mathbf{W}_1$. }
    \label{fig:orig_decoupling}
\end{figure}

Finding compact representations of multivariate functions forms an essential part of many scientific fields, such as block-structured system identification \cite{dreesen2016decoupling} and deep neural network compression \cite{zniyed2021tensor}. The decoupling methodology \cite{dreesen2015decoupling} is a powerful technique  that aims at representing a function by a composition of linear transformations, with elementwise nonlinear functions sandwiched between two matrices (see figure \ref{fig:orig_decoupling}). The advantage of the decoupling problem is that it allows to leverage a tensor decomposition, which makes it attractive in many applications ~\cite{zniyed2021tensor, hollander2017multivariate}. The decoupling representations can be viewed as neural networks with trainable activation functions, per neuron \cite{zniyed2021tensor}.

Dreesen et al. \cite{dreesen2015decoupling} introduced a tensor-based method for solving the decoupling problem based on first-order information of $\mathbf{f}(\mathbf{x})$ and the computation of the Canonical Polyadic Decomposition (CPD) of a third-order tensor. In the noiseless case it is guaranteed to solve the decoupling problem thanks to the uniqueness properties of the CPD. To deal with more practical noisy, or non-unique CPD scenarios, several approaches have been proposed, including ~\cite{hollander2017multivariate, zniyed2021tensor, zniyed2021learning, decuyper2022decoupling}. One of the key issues is that the nonlinear functions in figure \ref{fig:orig_decoupling} need to be estimated.

To this end, Hollander et al. \cite{hollander2017multivariate} propose parameterizing the internal functions $g_i$ as polynomials of a certain degree. Zniyed et al. \cite{zniyed2021tensor} provide a more general basis function representation but choose to parameterize the functions $g_i$ as piecewise-linear functions for their application of neural network compression. In contrast, Decuyper et al. \cite{decuyper2022decoupling} present an algorithm with a fully non-parametric representation of the decoupling's internal functions.

The drawback of existing approaches is that they mostly assume a simple parametric form of the internal functions (polynomial and piecewise linear) ~\cite{zniyed2021tensor, hollander2017multivariate}. The exception is Decuyper's work \cite{decuyper2022decoupling}, which presents a non-parametric 'filtering' approach to estimating internal functions, but their approach is mostly heuristic and comes at a higher computational cost.


In this paper, we propose a principled alternative that uses B-splines for the internal functions and develop an algorithm for it. Such a choice allows us to develop a coupled tensor factorization algorithm that is efficient, but can provide a non-parametric estimation of the internal functions. Unlike polynomials, this gives better behavior of the algorithm steps.

\section{Notations and tensor background}
\subsection{Notations}
Matrices and vectors are denoted with bold capital and lowercase letters, respectively. 
Tensors are denoted by calligraphic capital letters. 
For a third order tensor $\mathcal{X}$ of size $I \times J \times K$, the $i$-th horizontal, $j$-th lateral and $k$-th frontal slice are denoted by $\mathcal{X}_{i,:,:}$, $\mathcal{X}_{:,j,:}$ and $\mathcal{X}_{:,:,k}$ respectively. The operation $\operatorname{unfold}_k(\mathcal{X})$ unfolds the tensor $\mathcal{X}$ over its $k$-th mode as described in \cite{kolda2009tensor}. The $i$-th row and $j$-th column of a matrix $\mathbf{A}$ are denoted as $\mathbf{A}^{i,:}$ and $\mathbf{A}^{:,j}$ respectively. The symbol $\odot$ denotes the Khatri-Rao product. Finally, the $\operatorname{diag}(.)$ operation forms a diagonal matrix where the main diagonal is the vector that is provided as parameter and $\lVert . \rVert$ denotes the norm of a tensor, defined as the square root of the sum of the squares of its elements.

\subsection{Canonical polyadic decomposition}
The \textit{canonical polyadic decomposition} (CPD) \cite{kolda2009tensor} of a third-order tensor $\mathcal{X} \in \mathbb{R}^{I \times J \times K}$ expresses the tensor as a sum of rank-one tensors,
or alternatively,
the tensor $\mathcal{X}$ admits a CPD if its slices can be represented as 
\begin{equation}
    \mathcal{X}_{:,:,k} = \mathbf{A} \cdot \text{diag}(\mathbf{C}^{k,:}) \cdot \mathbf{B}, \text{ for } k=1,2,\hdots,K, \label{eq:CPD}
\end{equation}
where $\mathbf{A} \in \mathbb{R}^{I \times r}$, $\mathbf{B} \in \mathbb{R}^{r \times J}$ and $\mathbf{C} \in \mathbb{R}^{K \times r}$ 
are the 
factor matrices.
We use 
the notation  $\mathcal{X}  =[\![\mathbf{A}, \mathbf{B}, \mathbf{C}]\!]$.
The canonical rank is the smallest value $r$ for which equation \eqref{eq:CPD} holds. 

The CPD described in equation \eqref{eq:CPD} is unique under mild conditions. Uniqueness here means that  the CPD is unique up to the following scaling and permutation ambiguities \cite{kolda2009tensor}:
\begin{equation}
    \mathcal{X} = [\![\mathbf{A}\mathbf{\Pi}\mathbf{\Lambda_{A}}, \mathbf{B}\mathbf{\Pi}\mathbf{\Lambda_{B}}, \mathbf{C}\mathbf{\Pi}\mathbf{\Lambda_{C}}]\!], \nonumber
\end{equation}
with permutation matrix $\mathbf{\Pi} \in \mathbb{R}^{r \times r}$ and diagonal matrices $\mathbf{\Lambda_{A}}$, $\mathbf{\Lambda_{B}}$, $\mathbf{\Lambda_{C}}$ for which $\mathbf{\Lambda_{A}}\mathbf{\Lambda_{B}}\mathbf{\Lambda_{C}}=\mathbf{I}$.
Several sufficient uniqueness conditions exist (for example  Kruskal's condition), see  \cite{sidiropoulos2017tensor} for an overview.

\section{Decoupling}

\subsection{The decoupling problem}
Dreesen et al. \cite{dreesen2015decoupling} formulates the decoupling problem as follows: given a multivariate vector function $\mathbf{f}: \mathbb{R}^m \rightarrow \mathbb{R}^n$, find a decoupled representation of $\mathbf{f}(\mathbf{x})$ such that
\begin{equation}
    \mathbf{f}(\mathbf{x}) = \mathbf{W}_1 \mathbf{g}(\mathbf{W}_0 \mathbf{x}) \label{eq:decoupled_rep}
\end{equation}
with linear transformation matrices $\mathbf{W}_1 \in \mathbb{R}^{n \times r}, \mathbf{W}_0 \in \mathbb{R}^{r \times m}$ and $\mathbf{g}(\mathbf{u}) =
    \begin{bmatrix}
        g_1(u_1) & g_2(u_2) & \cdots & g_r(u_r)
    \end{bmatrix}^{\top} \in \mathbb{R}^r$ consists of univariate functions $g_i: \mathbb{R} \rightarrow \mathbb{R}$.
Dreesen et al. \cite{dreesen2015decoupling} proposes a tensor-based solution strategy that uses the first-order information of $\mathbf{f}(\mathbf{x})$. This first order information is encapsulated in the Jacobian $\mathbf{J}_{\mathbf{f}}(\mathbf{x})$
\begin{equation}
    \mathbf{J}_{\mathbf{f}}(\mathbf{x}) = \begin{bmatrix}
        \dfrac{\partial f_1(\mathbf{x})}{\partial x_1} & \hdots & \dfrac{\partial f_1(\mathbf{x})}{\partial x_m} \\
        \vdots & & \vdots \\
        \dfrac{\partial f_n(\mathbf{x})}{\partial x_1} & \hdots & \dfrac{\partial f_n(\mathbf{x})}{\partial x_m}        
    \end{bmatrix} \in \mathbb{R}^{n \times m}, \nonumber
\end{equation}
where, under the assumption that $\mathbf{f}(\mathbf{x})$ follows the model in equation \eqref{eq:decoupled_rep}, it holds that
\begin{equation}
    \mathbf{J}_{\mathbf{f}}(\mathbf{x}) = \mathbf{W}_1 \; \text{diag}(g'_i(\mathbf{W}^{i,:}_0 \mathbf{x}) \; \mathbf{W}_0. \nonumber
\end{equation}
Next, $\mathbf{J}_{\mathbf{f}}(\mathbf{x})$ can be evaluated in $S$ sample points $\mathbf{x}^{(s)} \in \mathbb{R}^s$, for $s=1,2,\hdots,S$
\begin{align}
    \mathbf{J}_{\mathbf{f}}(\mathbf{x}^{(s)}) &= \mathbf{W}_1 \; \text{diag}(g'_i(\mathbf{W}^{i,:}_0 \mathbf{x}^{(s)}) \; \mathbf{W}_0 = \mathbf{W}_1 \; \mathbf{D}_{\mathbf{g}}^{(s)} \; \mathbf{W}_0, \nonumber
\end{align}
note here that 1) $\mathbf{D}_{\mathbf{g}}^{(s)} \in \mathbb{R}^{r \times r}$ is diagonal and 2) $\mathbf{W}_0$ and $\mathbf{W}_1$ are independent of the sample point $\mathbf{x}^{(s)}$ \cite{dreesen2015decoupling}.

As a result, stacking the Jacobian matrices $\mathbf{J}_{\mathbf{f}}(\mathbf{x}^{(s)})$, for $s=1,2,\hdots,S$  as frontal slices of a third-order tensor yields a Jacobian tensor $\mathcal{J} \in \mathbb{R}^{n \times m \times S}$ for which
\begin{equation}
    \mathcal{J}_{:,:,s} = \mathbf{J}_{\mathbf{f}}(\mathbf{x}^{(s)}) = \mathbf{W}_1 \; \mathbf{D}^{(s)}_{\mathbf{g}} \; \mathbf{W}_0, \label{eq:frontal_slice_J}
\end{equation}
for $s=1,2,\hdots,S$. This shows that by construction, the tensor $\mathcal{J}$ admits a CPD $\mathcal{J} = [\![\mathbf{W}_1, \mathbf{W}^{\top}_0, \mathbf{G}]\!]$
where for $\mathbf{W}_0\mathbf{x}^{(s)} = \mathbf{u}^{(s)} \in \mathbb{R}^r$, for $s=1,2,\hdots,S$, the factor matrix $\mathbf{G}$ is 
\vspace*{-1mm}
\begin{align}
    \mathbf{G} =  \begin{bmatrix}
        g'_1(u^{(1)}_1) & \hdots & g'_r(u^{(1)}_r) \\
        \vdots & & \vdots \\
        g'_1(u^{(S)}_1) & \hdots & g'_r(u^{(S)}_r) \\
    \end{bmatrix}. \label{eq:structure_G}
\end{align}
Equation \eqref{eq:frontal_slice_J} indicates that computing the CPD of $\mathcal{J}$ yields the factor matrices $\mathbf{W}_1$ and $\mathbf{W}_0$ of the decoupled model \eqref{eq:decoupled_rep} as well as the matrix $\mathbf{G} \in \mathbb{R}^{S \times r}$ which contains first-order information of the internal functions $g_i$, for $i=1,2,\hdots,r$.

The tensor-based solution strategy 
is summarized as:
\begin{enumerate}
    \item Evaluate the Jacobian of $\mathbf{f}(\mathbf{x})$ in S sample points $\mathbf{x}^{(s)}$, yielding $\mathbf{J}_{\mathbf{f}}(\mathbf{x}^{(1)}), \mathbf{J}_{\mathbf{f}}(\mathbf{x}^{(2)}), \hdots,\mathbf{J}_{\mathbf{f}}(\mathbf{x}^{(S)})$.
    \item Stack the Jacobian matrices $\mathbf{J}_{\mathbf{f}}(\mathbf{x}^{(s)})$, for $s=1,2,\hdots,S$, into the tensor $\mathcal{J} \in \mathbb{R}^{n \times m \times S}$.
    \item Compute the CPD of $\mathcal{J}$, yielding $\mathbf{W}_1, \mathbf{W}_0$ and $\mathbf{G}$ up to scaling and permutation ambiguities.
    \item Use the first-order information in $\mathbf{G}$ to determine the representation of the decoupling's internal functions $g_i$.
\end{enumerate}

\subsection{Representation of the internal functions} 

Ideally, a representation is chosen such that the decoupling's internal functions can be represented in a non-parametric way without adding substantial algorithmic complexity. To this end, this work introduces the use of B-spline functions to represent the internal functions $g_i$ (or derivatives $g'_i$), for $i=1,2,\hdots,r$,
\vspace{-15px}
\begin{gather}
    g_i(u_i) = c_{i,0} + \sum^{df}_{j=1} c_{i,j} B^{\mathbf{\Delta}_i}_{j,d}(u_i), \text{ derivative to get $g'_i$} \label{eq:B-spline_g}, \\[-0.5em]
    \text{or} \nonumber \\[-0.5em]
    g'_i(u_i) = \sum^{df}_{j=1} c_{i,j} B^{\mathbf{\Delta}_i}_{j,d-1}(u_i), \text{ integrate to get $g_i$}, \label{eq:B-spline_g_prime}
\end{gather}
\vspace{-2px}
where $df$, $d$ and $\mathbf{\Delta}_i$ are the degrees of freedom (DoF), order and knot vector of the spline respectively. The coefficients $c_{i,j}$ are to be learned as part of the decoupling problem.

The proposed B-spline representation allows to easily incorporate constraints on the internal functions $g_i$, such as nonnegativity or monotonicity, by constraining the coefficients of the spline to be nonnegative. This is not possible for a polynomial parameterization or the non-parametric method of Decuyper et al. \cite{decuyper2022decoupling}. 

\subsection{Optimization problem}

A critical part of the tensor-based solution strategy is the computation of the CPD of $\mathcal{J}$. The following optimization problem formulates the unstructured CPD of $\mathcal{J}$
\begin{equation}
    \displaystyle{\min_{\mathbf{W}_1, \mathbf{W}_0, \mathbf{G}}} \;\;  \lVert \mathcal{J} - [\![\mathbf{W}_1, \mathbf{W}^{\top}_0, \mathbf{G}]\!] \rVert^2. \label{eq:unstructured_CPD}
\end{equation}
However, optimization problem \eqref{eq:unstructured_CPD} only uses first-order information of $\mathbf{f}(\mathbf{x})$. Because of this, it is unable to approximate the constant terms of the internal functions $g_i$ and the resulting system $\widehat{\mathbf{f}}(\mathbf{x})$ will show a bias relative to the actual system $\mathbf{f}(\mathbf{x})$. The two main strategies to solve this problem are 1) correcting the bias of the computed decoupling $\widehat{\mathbf{f}}(\mathbf{x})$ as a whole in a second step, as done by Decuyper et al. \cite{decuyper2022decoupling}, and 2) adding zeroth-order information into the optimization problem to directly estimate the constant terms of the internal functions $g_i$, as done by Zniyed et al. \cite{zniyed2021tensor}.

This work adopts the strategy of \cite{zniyed2021tensor} and integrates a zeroth-order information matrix $\mathbf{F} \in \mathbb{R}^{n \times S}$ into optimization problem \eqref{eq:unstructured_CPD}. The structure of $\mathbf{F}$ is given by
\begin{align}
    \mathbf{F} &= \begin{bmatrix}
        \mathbf{f}(\mathbf{x}^{(1)}) & \mathbf{f}(\mathbf{x}^{(2)}) & \cdots & \mathbf{f}(\mathbf{x}^{(S)})
    \end{bmatrix}\nonumber \\ 
    &= \mathbf{W}_1 \begin{bmatrix}
        \mathbf{g}(\mathbf{u}^{(1)}) & \mathbf{g}(\mathbf{u}^{(2)}) & \cdots & \mathbf{g}(\mathbf{u}^{(S)})
    \end{bmatrix} \label{eq:structure_R} \\
    &=\mathbf{W}_1 \cdot \mathbf{R}^{\top}, \nonumber
\end{align}
where $\mathbf{u}^{(s)} = \mathbf{W}_0 \mathbf{x}^{(s)} \in \mathbb{R}^r$ and $\mathbf{R} \in \mathbb{R}^{S \times r}$ contains zeroth-order information of the internal functions $g_i$. 

Incorporating $\mathbf{F}$ into optimization problem \eqref{eq:unstructured_CPD} results in a coupled matrix-tensor factorization (CMTF) ~\cite{liu2021tensors, zniyed2021tensor}
\begin{align}
    \!\!\displaystyle{\min_{\mathbf{W}_1, \mathbf{W}_0, \mathbf{G}, \mathbf{R}}} &\;\;  \!\!\lVert \mathcal{J} - [\![\mathbf{W}_1, \mathbf{W}^{\top}_0\!, \mathbf{G}]\!] \rVert^2 + \lambda \lVert \mathbf{F} - \mathbf{W}_1 \mathbf{R}^{\top}\! \rVert^2. \!\label{eq:unstructured_CPD_2}
\end{align}
%
This problem 
still computes an unstructured CPD of $\mathcal{J}$. However if the reconstruction error of the CPD is non-zero, or the computed CPD is not unique then there is no guarantee that the computed $\mathbf{G}$ and $\mathbf{R}$ have the required structure.

We take this problem into account by forcing the required structure onto $\mathbf{G}$ and $\mathbf{R}$ through the B-spline representation of the internal functions given in equations \eqref{eq:B-spline_g} and \eqref{eq:B-spline_g_prime}. Equations \eqref{eq:structure_G} and \eqref{eq:structure_R} show the structure of $\mathbf{G}$ and $\mathbf{R}$ respectively. This structure is enforced by adding the following constraints on the columns of $\mathbf{G}$ and $\mathbf{R}$
\begin{align}
    \mathbf{G}^{:,j} = \mathbf{B}_j \cdot \mathbf{c}_{j} \;\;\;\; \text{for } j = 1,2,\hdots,r, \label{eq:constraint_G} \\
    \mathbf{R}^{:,j} = \widetilde{\mathbf{B}}_j \cdot \mathbf{c}_{j} \;\;\;\; \text{for } j = 1,2,\hdots,r, \label{eq:constraint_R}
\end{align}
\noindent
where $\mathbf{c}_j = \begin{bmatrix}
    c_{j,0} & c_{j,1} & \cdots & c_{j,df}
\end{bmatrix}^{\top} \in \mathbb{R}^{df + 1}$. The matrices $\mathbf{B}_j \in \mathbb{R}^{S \times df + 1}$ and $\widetilde{\mathbf{B}}_j \in \mathbb{R}^{S \times df + 1}$ are the evaluated B-spline design matrices at the values $u^{(s)}_j = (\mathbf{W}_0 \mathbf{x}^{(s)})_j$, for $s=1,2,\hdots,S$, with an extra first column such that $\mathbf{B}^{:,0}$ is a column of zeros and $\widetilde{\mathbf{B}}^{:,0}$ is a column of ones. The extra column is to take into account the constant term $c_{j,0}$ in $\mathbf{c}_j$.

Incorporating \eqref{eq:constraint_G} and \eqref{eq:constraint_R} as constraints in optimization problem \eqref{eq:unstructured_CPD_2} yields the final optimization problem
\begin{align}
    \!\!\displaystyle{\min_{\mathbf{W}_1, \mathbf{W}_0, \mathbf{G}, \mathbf{R}}} &\;\;  \!\!\lVert \mathcal{J} - [\![\mathbf{W}_1, \mathbf{W}^{\top}_0\!, \mathbf{G}]\!] \rVert^2 + \lambda \lVert \mathbf{F} - \mathbf{W}_1 \mathbf{R}^{\top} \rVert^2 \label{eq:final_optimization}\\
     \operatorname{s.t.} \quad &\mathbf{G}^{:,j} = \mathbf{B}_j \cdot \mathbf{c}_{j} \;\;\;\; \text{for } j = 1,2,\hdots,r \nonumber \\
     &\mathbf{R}^{:,j} = \widetilde{\mathbf{B}}_j \cdot \mathbf{c}_{j} \;\;\;\; \text{for } j = 1,2,\hdots,r. \nonumber
\end{align}
The $\lambda$ parameter is typically set to a fixed low value, $0.01$ or $0.1$, and stays fixed during execution or is increased over time.

\section{Algorithm}

This work uses the efficient projection strategy algorithm introduced by Zniyed et al. \cite{zniyed2021tensor} to solve optimization problem \eqref{eq:final_optimization}. But, the projection step is adapted to facilitate the B-spline representations \eqref{eq:B-spline_g} and \eqref{eq:B-spline_g_prime} and allow nonnegativity or monotonicity constraints on the internal functions. The use of B-splines combines the efficiency of the projection algorithm with the non-parametric property of Decuyper's algorithm \cite{decuyper2022decoupling}.

Algorithm \ref{alg:CMTF-BSD} shows the full algorithm, called CMTF-BSD (\textbf{CMTF} \textbf{B}-\textbf{S}pline \textbf{D}ecoupling). The CMTF-BSD algorithm normalizes the columns of $\mathbf{W}^{\top}_0$ for improved conditioning, which is not part of the algorithm proposed by Zniyed \cite{zniyed2021tensor}. The normalization procedure is given in algorithm \ref{alg:normalize_columns}.

\begin{algorithm}[h!]
    \caption{CMTF-BSD algorithm}
    \label{alg:CMTF-BSD}
    \begin{algorithmic}[1]
        \Require $\mathcal{J} \in \mathbb{R}^{n \times m \times S}$, $\mathbf{F} \in \mathbb{R}^{n \times S}$, $df$, $d$, $r$, samples,
        \State $\mathbf{W}_0,  \mathbf{G}, \mathbf{R} \gets$ Random initialization
        \While{stop criteria not met}
            \State $\mathbf{W}_1 \gets \displaystyle{\argmin_{\mathbf{W}_1}} \lVert \text{unfold}_1(\mathcal{J}) - \mathbf{W}_1 ( \mathbf{G} \odot \mathbf{W}^{\top}_0)^{\top} \rVert^2 \nonumber$
            \vspace{-2px}
            \State $\quad\quad\quad\quad\quad\quad\quad\quad\quad\quad\quad\quad\quad+ \lambda \lVert \mathbf{F} - \mathbf{W}_1 \mathbf{R}^{\top} \rVert^2$
            \vspace{2px}
            \State $\mathbf{W}_0 \gets \displaystyle{\argmin_{\mathbf{W}_0}} \lVert \text{unfold}_2(\mathcal{J}) - \mathbf{W}^{\top}_0 ( \mathbf{G} \odot \mathbf{W}_1)^{\top} \rVert^2$
            \vspace{2px}
            \State
            $\mathbf{W}_0, \mathbf{W}_1 \gets \text{Normalize}\_\text{columns}\_\mathbf{W}^{\top}_0(\mathbf{W}_0, \mathbf{W}_1)$
            \vspace{4px}
            \State $\mathbf{G} \gets \displaystyle{\argmin_{\mathbf{G}}} \lVert \text{unfold}_3(\mathcal{J}) - \mathbf{G} (\mathbf{W}^{\top}_0 \odot \mathbf{W}_1)^{\top} \rVert^2$
            \State $\mathbf{R} \gets \displaystyle{\argmin_{\mathbf{R}}} \lVert \mathbf{F} - \mathbf{W}_1 \mathbf{R}^{\top} \rVert^2$
            \State $\text{xSamples} \gets \mathbf{W}_0 \cdot$ samples
            \State $\mathbf{G}, \mathbf{R} \gets \text{Bspline\_projection}\left(\mathbf{G}, \mathbf{R}, df, d, \text{xSamples}\right)$
        \EndWhile
    \Ensure $\mathbf{W}_1, \mathbf{W}_0, \mathbf{G}, \mathbf{R}$
    \end{algorithmic}
\end{algorithm}

\begin{algorithm}[h!]
    \caption{Normalize\_columns\_$\mathbf{W}^{\top}_0$}
    \label{alg:normalize_columns}
    \begin{algorithmic}[1]
        \Require $\mathbf{W}_0$, $\mathbf{W}_1$
        \For{$i=1,2,\hdots,r$}
            \State $\beta \gets \lVert (\mathbf{W}^{\top}_0)^{:,i} \rVert$
            \vspace{2px}
            \State $(\mathbf{W}^{\top}_0)^{:,i} \gets (\mathbf{W}^{\top}_0)^{:,i} / \beta$
            \vspace{2px}
            \State $(\mathbf{W}_1)^{:,i} \gets \beta \; (\mathbf{W}_1)^{:,i} $
        \EndFor
    \Ensure $\mathbf{W}_0, \mathbf{W}_1$
    \end{algorithmic}
\end{algorithm}

\begin{algorithm}[h!]
    \caption{Bspline\_projection}
    \label{alg:Bspline_projection}
    \begin{algorithmic}[1]
        \Require $\mathbf{G}$, $\mathbf{R}$, $df$, $d$, xSamples $\in \mathbb{R}^{r \times S}$
        \vspace{5px}
        \For{$j=1,2,\hdots,r$}
            \State $\mathbf{x}_j \gets $ xSamples$^{j,:}$
            \State $\mathbf{\Delta}_j \gets \text{Determine\_knots}(\mathbf{x}_j, df, d)$
            \If{Representation \eqref{eq:B-spline_g} used}
                \State $\widetilde{\mathbf{B}} \gets \text{Design\_matrix}(\mathbf{x}_j, \mathbf{\Delta}_j, df, d)$
                \State $\mathbf{B} \gets \text{Derivative}(\widetilde{\mathbf{B}})$
            \ElsIf{Representation \eqref{eq:B-spline_g_prime} used}
                \State $\mathbf{B} \gets \text{Design\_matrix}(\mathbf{x}_j, \mathbf{\Delta}_j, df, d)$          \State $\widetilde{\mathbf{B}} \gets \text{Integrate}(\mathbf{B})$
            \EndIf
            \State $\mathbf{c}_j \gets \displaystyle{\argmin_{\mathbf{c}_j}} \lVert \mathbf{G}^{:,j} - \mathbf{B} \; \mathbf{c}_j \rVert^2 + \lambda \lVert \mathbf{R}^{:,j} - \widetilde{\mathbf{B}} \; \mathbf{c}_j \rVert^2$
            \State $\mathbf{G}^{:,j} \gets \mathbf{B} \; \mathbf{c}_j$, $\mathbf{R}^{:,j} \gets \widetilde{\mathbf{B}} \; \mathbf{c}_j$
        \EndFor
        \vspace{5px}
        \Ensure $\mathbf{G}, \mathbf{R}$
    \end{algorithmic}
\end{algorithm}

Algorithm \ref{alg:Bspline_projection} shows the B-spline projection step (line $10$ in algorithm \ref{alg:CMTF-BSD}) in more detail. The Determine\_knots function retrieves the knots from the input values $\mathbf{x}_i$ based on quantiles, which depend on the degrees of freedom $df$ and the order $d$ of the spline \cite{deboor1978splines}. The Design\_matrix(.) function constructs the B-spline design matrix for the given parameters $\mathbf{\Delta}_j, df$ and $d$, evaluated at the points $\mathbf{x}_j$. The Derivative(.) and Integrate(.) functions construct the B-spline design matrix for the derivatives and indefinite integrals respectively, of the B-spline basis that was used to construct the design matrix that is given as a parameter, evaluated in the same points.

Enforcing a monotonicity or nonnegativity constraint on the internal functions $g_i$ can be done by using nonnegative least squares to solve for $\mathbf{c}_j$ on line $11$ in algorithm \ref{alg:Bspline_projection}. 

\section{Experiments and results}

\subsection{Metrics}
The following metrics are used in the experiments
\begin{gather}
    \text{Error}(\mathcal{J}) = \dfrac{\lVert \mathcal{J} - \widehat{\mathcal{J}} \rVert^2}{\lVert \mathcal{J} \rVert^2}, \nonumber \\
    e_i = \dfrac{\sqrt{\frac{1}{S}\sum^S_{s=1}\left(f_i(\mathbf{x}^{(s)}) - \widehat{f}_i(\mathbf{x}^{(s)})\right)^2}}{\sqrt{\frac{1}{S}\sum^S_{s=1}\left(f_i(\mathbf{x}^{(s)}) - \mathbb{E}(f_i)\right)^2}} \times 100, \nonumber
\end{gather}
where $\widehat{\mathcal{J}}$ 
is the approximation of $\mathcal{J}$ 
resulting from the CMTF-BSD algorithm. Note here that $e_i$ is a relative error on the $i$th output of $\mathbf{f}(\mathbf{x})$, as a percentage. 

\subsection{Effect of DoF and degree of spline representation}\label{subsec:flex_exp}

In this section, the CMTF-BSD(.) algorithm is executed on a system with trigonometric internal functions
\vspace{2px}
\begin{gather}
    \mathbf{f}_{trig}(\mathbf{x}) = \mathbf{W}_1\mathbf{g}(\mathbf{W}_0\mathbf{x}), \label{eq:sys_trig}
\end{gather}
\vspace{-20px}
\begin{gather}
    \mathbf{W}_1 = \begin{bmatrix}
        -1.7 & -2.3 & 2.5 \\
        0.5 & -0.5 & 0.2
    \end{bmatrix}, \mathbf{W}_0 = \begin{bmatrix}
        2.1 & -1 \\
        0.4 & -1.8 \\
        -1.6 & -0.2
    \end{bmatrix}, \nonumber
\end{gather}
\vspace{-5px}
\begin{gather}
    \mathbf{g}(\mathbf{u}) = \begin{bmatrix}
        sin(u_1) + 2 \\
        cos(2 \cdot u_2) - 1.5 \\
        sin(2 \cdot u_3) + \dfrac{u_3}{2}\\
    \end{bmatrix} \nonumber
\end{gather}

The goal of the CMTF-BSD(.) algorithm is to retrieve the decoupled representation \eqref{eq:sys_trig} from the Jacobian tensor $\mathcal{J}$ and zeroth-order information matrix $\mathbf{F}$. The decoupled representation contains $3$ internal functions and only $2$ inputs and outputs, making the associated CPD of $\mathcal{J}$ not unique.

For the experiment, the algorithm is executed $30$ times for different spline degrees ($d=1,2,3$) and DoF ($4,6,\hdots,28$), using representation \eqref{eq:B-spline_g}. After executing the CMTF-BSD algorithm, the discovered internal function shapes are interpolated by polynomials of degree $10$. The Jacobian is constructed with $S=100$ sample points drawn from $\mathcal{U}(-1.5,1.5)$, yielding the tensor $\mathcal{J} \in \mathbb{R}^{2 \times 2 \times 100}$ and matrix $\mathbf{F} \in \mathbb{R}^{2 \times 100}$. Each execution draws a new set of $100$ sample points.

\begin{figure}
    \centering
     \includegraphics[width=0.55\textwidth]{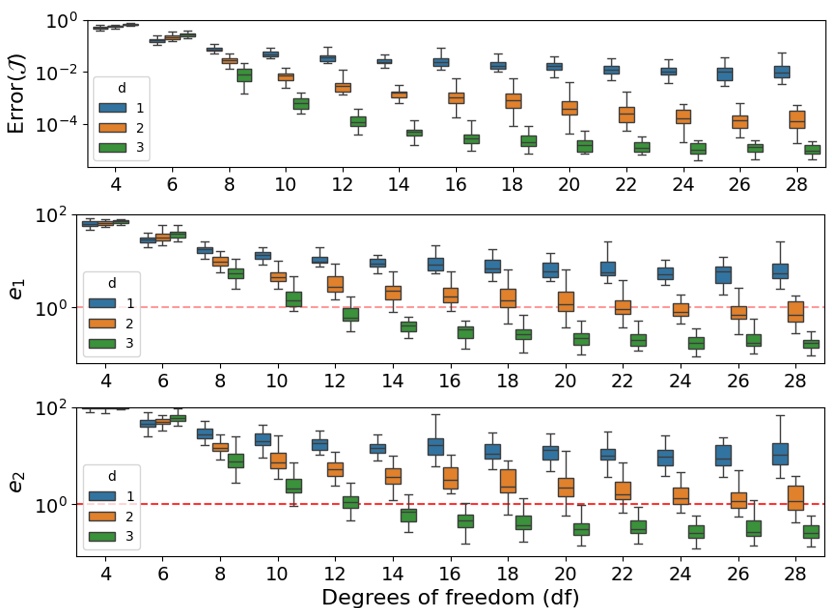}
    \caption{Results for applying the CMTF-BSD(.) algorithm \ref{alg:CMTF-BSD} to the sytem $\mathbf{f}_{trig}$ of equation \eqref{eq:sys_trig}, for $30$ executions per $(d,df)$ pair where $d \in \{1,2,3\}$ and $df \in \{4,6,\hdots,28\}$. Top figure: the relative reconstruction error of the Jacobian tensor $\mathcal{J}$; middle figure: the relative error of the computed system for the first output; bottom figure: the relative error for the second output. The red dotted line indicates an error of $1\%$.}
    \label{fig:f_trig_results}
\end{figure}

Figure \ref{fig:f_trig_results} shows the results for $\mathbf{f}_{trig}$ (outliers are not plotted for readability). The cubic spline representation, i.e., $d=3$, performs the best yielding output errors below $1\%$ when $df \geq 12$. The approximation of $\mathbf{f}_{trig}$ is worse for $d=2$, as less flexibility requires a higher $df$ value to reach output errors below $1\%$. The case for which $d=1$ gives the worst results. The executions for which the output errors drop below $1\%$ are able (or close) to recover the system $\mathbf{f}_{trig}$, even though the CPD itself is not unique.  

The problem of recovering $\mathbf{f}_{trig}$ can be seen as a regression task for which the case of $d=1$ is not optimal since this results in piece-wise linear functions $g_i$ that are interpolated by polynomials in a second step. For the piece-wise linear case of $d=1$, a neural network or decision boundary problem such as in \cite{zniyed2021tensor} is more relevant. Do note here that compared to \cite{zniyed2021tensor}, achieving piece-wise linear functions does not require a different basis as we just set the degree of the B-spline representation equal to $1$.


\subsection{Monotonically increasing internal functions}\label{subsec:mono_exp}

This section explores the incorporation of a monotonicity constraint on the decoupling's internal functions $g_i$. The constraint is enforced through the use of the B-spline representation \eqref{eq:B-spline_g_prime} where the coefficients $\mathbf{c}_j$ on line $11$ of the Bspline\_projection algorithm \ref{alg:Bspline_projection} are computed using nonnegative least squares. Since B-spline basis functions are nonnegative, this results in nonnegative derivatives $g'_i$ and monotonically increasing internal functions $g_i$.

However, it is possible when using nonnegative least squares that during intermediate steps of the algorithm the coefficients $\mathbf{c}_j$ are computed to be all zero, which crashes the algorithm if not taken into account. To take this into account, an extra check is added after the computation of the coefficients $\mathbf{c}_j$ where, if the computed coefficients are all zero, the function $g'_i$ is replaced by a monotonically increasing LeakyReLU activation function \cite{lederer2021activation} with negative slope $-0.5$.

The system used for the experiment has $3$ inputs, $3$ outputs and $3$ monotonically increasing internal functions\\ 
\hspace*{2mm} $\mathbf{g}(\mathbf{u}) = \begin{bmatrix}
    g_1(u_1) & g_2(u_2) & g_3(u_3)
\end{bmatrix}^{\top}$,
\begin{gather}
    g_1(u_1) = \dfrac{u^3_1}{3} + u_1, \;\; g_2(u_2) = e^{u_2}, \;\; g_3(u_3) = \dfrac{1}{1-e^{-u_3}}. \nonumber
\end{gather}
The entries of the factor matrices $\mathbf{W}_0 \in \mathbb{R}^{3 \times 3}$ and $\mathbf{W}_1 \in \mathbb{R}^{3 \times 3}$ are randomly drawn from $\mathcal{U}(-2,2)$, for each execution of the algorithm.

\begin{figure}
    \centering
    \includegraphics[width=0.75\linewidth]{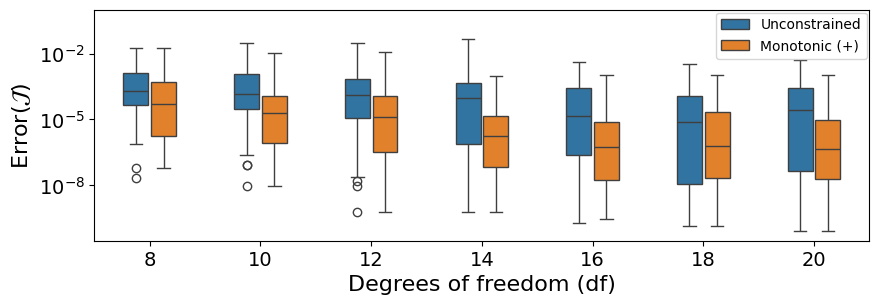}
    \caption{Reconstruction errors for the Jacobian tensor $\mathcal{J}$, for $30$ executions of the CMTF-BSD algorithm \ref{alg:CMTF-BSD} and degrees of freedom $df \in \{8,10,12,\hdots,20\}$, with and without monotonicity constraint. The monotonic (+) results compute the coefficients on line $11$ of algorithm \ref{alg:Bspline_projection} with nonnegative least squares to retrieve monotonically increasing functions.}
    \label{fig:results_mono}
\end{figure}

Figure \ref{fig:results_mono} shows the results for $30$ executions of the CMTF-BSD algorithm for the described system with $d=4$ and $df \in \{8,10,12,\hdots,20\}$. The algorithm is executed both with and without the monotonicity constraint, denoted as 'Monotonic (+)' and 'Unconstrained' respectively. 

Figure \ref{fig:results_mono} shows that the reconstruction errors for $\mathcal{J}$ are improved by adding the monotonicity constraint. Furthermore table \ref{tab:table_monotone} shows that for the unconstrained case, even if the underlying functions of the system are monotonically increasing or decreasing, there is no guarantee that the resulting functions are monotonic. On the other hand, by enforcing the monotonicity constraint the resulting decouplings from all $30$ executions have certified monotonically increasing internal functions.
\vspace{10px}
\begin{table}[htbp]
    \caption{Number of executions of the CMTF-BSD algorithm (out of $30$) for $df \in \{8,10,12,\hdots,20\}$, where the resulting decoupling has certified monotonic internal functions.}
    \begin{center}
    \begin{tabular}{|c||c|c|c|c|c|c|c|}
    \hline
    & \multicolumn{7}{|c|}{\textbf{Degrees of freedom} ($df$)} \\\cline{2-8}
     & $\mathbf{8}$ & $\mathbf{10}$ & $\mathbf{12}$ & $\mathbf{14}$ & $\mathbf{16}$ & $\mathbf{18}$ & $\mathbf{20}$ \\
    \hline
    \hline
    \textbf{Unconstrained} & $7$ & $7$ & $7$ & $9$ & $15$ & $16$ & $13$ \\
    \hline
    \textbf{Monotonic (+)} &  \textbf{30} & \textbf{30} & \textbf{30} & \textbf{30} & \textbf{30} & \textbf{30} & \textbf{30} \\
    \hline
    \end{tabular}
    \label{tab:table_monotone}
    \end{center}
\end{table}
\vspace{-10px}
\section{Conclusion}

This work introduced the CMTF-BSD decoupling algorithm, which combines the efficiency of the decoupling algorithm of Zniyed \cite{zniyed2021tensor} and the non-parametric property of that of Decuyper \cite{decuyper2022decoupling}. This is done by 1) using the projection algorithm of \cite{zniyed2021tensor} and 2) incorporating a B-spline representation of the internal functions to allow non-parametric estimation. In addition, as shown in the experiments, the use of B-splines allows to enforce nonnegativity and monotonicity constraints, which is not possible for a polynomial basis or the non-parametric method of Decuyper \cite{decuyper2022decoupling}. The constrained example in section \ref{subsec:mono_exp} shows good results for the incorporation of a monotonicity constraint and the fact that, even when the underlying functions are monotonic, adding the constraint is necessary for guaranteed monotonicity. Furthermore, the example in section \ref{subsec:flex_exp} shows that the CMTF-BSD algorithm performs well on a system with trigonometric functions, which would require a high degree polynomial to approximate. The CMTF-BSD algorithm achieves stable output errors below $1\%$ for different configurations of $d$ and $df$ and is able to recover the underlying system despite a non-unique CPD.

Future work includes analyzing the use of other spline types such as smoothing splines \cite{deboor1978splines} and natural cubic splines \cite{deboor1978splines}, improving the incorporation of constraints, providing a formal complexity analysis of the algorithm to compare with Decuyper \cite{decuyper2022decoupling}, applying the CMTF-BSD algorithm to different applications in system identification and neural network compression and formalizing the theoretical basis of the proposed algorithm through existing B-spline approximation theory. 

\section*{Acknowledgment}

This work was supported by the FWO (FWO Vlaanderen) fundamental research fellowship 11A2H25N.

\end{document}